\documentclass{article}
\usepackage{amsfonts}
\usepackage{amssymb}
\usepackage{amscd}

\usepackage{stmaryrd}

\newcommand{\ar}{{\mathfrak r}} 

\newcommand{\pfont}{\sffamily\bfseries}
\newcommand{\save}[1]{}

\renewcommand{\leq}{\leqslant}
\renewcommand{\geq}{\geqslant}

\newcommand{\li}{\item\zero}

\newcommand{\xfig}[2]{\medskip \centerline{\epsfig{figure=#1.eps,height=#2}} \bigskip}

\newcommand{\xfigtwo}[4]{\noindent \zero \hfill \epsfig{figure=#1.eps,height=#2} \onecm \epsfig{figure=#3.eps,height=#4} \hfill \zero}

\newcommand{\dara}[2]%
      {\overset{#1}{\underset{\raisebox{4ex}{\;\;\fnit{#2}}}{\longrightarrow}}} 

\newcommand{\myred}[1]{\Red{#1}}

\newcommand{\mybrown}[1]{\Mahogany{#1}}

\newcommand{\mywhite}[1]{\White{#1}}

\newcommand{\outt}[1]{}
\newcommand{\delete}[1]{}
\newcommand{\blank}[1]{}

\newcommand{\notedomission}[1]{\medskip\noindent{\bf TEXT OMITTED}\\[2mm]}

\newenvironment{xtr}{\huge\tt\hspace{-5mm}}{\\[2mm] \zero\hrulefill} 
\newcommand{\bxtr}{\begin{xtr}\begin{envviolet}\noindent}
\newcommand{\extr}{\end{envviolet}\end{xtr}}

\newcommand{\bmini}{\begin{minipage}[t]{0.8\textwidth}}
\newcommand{\emini}{\end{minipage}}

\newenvironment{envviolet}{\textViolet}{\textBlack}

\newcommand{\bxrm}[1]{\hbox{\rm #1}}

\newcommand{\bxbf}[1]{\hbox{\bf #1}}

\newcommand{\bxtt}[1]{\hbox{$\tt #1$}}

\newcommand{\fnit}[1]{\hbox{\footnotesize\it #1}}

\newcommand{\scriptbf}[1]{\hbox{\scriptsize\bf #1}}

\newcommand{\scriptbff}{\hbox{\scriptbf{f}}}

\newcommand{\scriptbfn}{\hbox{\scriptbf{n}}}

\newcommand{\scriptbfq}{\hbox{\scriptbf{q}}}

\newcommand{\bfb}{\hbox{\bf b}}
\newcommand{\bfc}{\hbox{\bf c}}

\newcommand{\bff}{\hbox{\bf f}}
\newcommand{\bfg}{\hbox{\bf g}}

\newcommand{\bfn}{\hbox{\bf n}}

\newcommand{\bfq}{\hbox{\bf q}}

\newcommand{\bft}{\hbox{\bf t}}

\newcommand{\tta}{\hbox{$\tt a$}}
\newcommand{\ttb}{\hbox{$\tt b$}}
\newcommand{\ttc}{\hbox{$\tt c$}}
\newcommand{\ttd}{\hbox{$\tt d$}}
\newcommand{\tte}{\hbox{$\tt e$}}
\newcommand{\ttf}{\hbox{$\tt f$}}
\newcommand{\ttg}{\hbox{$\tt g$}}

\newcommand{\ttl}{\hbox{$\tt l$}}

\newcommand{\ttp}{\hbox{$\tt p$}}
\newcommand{\ttq}{\hbox{$\tt q$}}

\newcommand{\tts}{\hbox{$\tt s$}}

\newcommand{\ttu}{\hbox{$\tt u$}}

\newcommand{\ttz}{\hbox{$\tt z$}}

\newcommand{\ttzero}{\hbox{$\tt 0$}}

\newcommand{\ttone}{\hbox{$\tt 1$}}

\newcommand{\gothc}{\mbox{$\mathfrak c$}}

\newcommand{\dN}{\hbox{$\mathbb N$}}

\newcommand{\grd}{\hbox{$\delta$}}
\newcommand{\gre}{\hbox{$\varepsilon$}}

\newcommand{\grl}{\hbox{$\lambda$}}

\newcommand{\grn}{\hbox{$\nu$}}
\newcommand{\grx}{\hbox{$\xi$}}

\newcommand{\grr}{\hbox{$\rho$}}
\newcommand{\grs}{\hbox{$\sigma$}}
\newcommand{\grt}{\hbox{$\tau$}}

\newcommand{\grf}{\hbox{$\varphi$}}

\newcommand{\grw}{\hbox{$\omega$}}

\newcommand{\grF}{\hbox{$\Phi$}}

\newcommand{\grW}{\hbox{$\Omega$}}

\newcommand{\bfgrw}{\hbox{$\pmb \omega$}}

\newcommand{\bgrw}{\hbox{\boldmath$\omega$}}

\newcommand{\sbcup}{\hbox{$\bigcup$}}

\newcommand{\ra}{\rightarrow}
\newcommand{\pa}{\rightharpoonup}
\newcommand{\sra}{\!\rightarrow\!} 

\newcommand{\rA}{\Rightarrow}  

\newcommand{\sminus}{\!-\!}

\newcommand{\suparrow}{\neghalfmm\uparrow\neghalfmm}

\newcommand{\sdownarrow}{\neghalfmm\downarrow\neghalfmm}
\newcommand{\sDownarrow}{\neghalfmm\Downarrow\neghalfmm}
\newcommand{\ssDownarrow}{\negonemm\Downarrow\negonemm}

\newcommand{\qed}{\hfill {\boldmath$\Box$}\\}
\newcommand{\lng}{\langle}
\newcommand{\rng}{\rangle}
\newcommand{\df}{=_{\rm df}}

\newcommand{\dfr}{\;\equiv_{\rm df}\;}

\newcommand{\ignore}[1]{}

\newcommand{\mx}{\makebox}

\newcommand{\zero}{\rule{0mm}{3mm}}

\newcommand{\negonemm}{\mx[-1mm]{}}
\newcommand{\neghalfmm}{\mx[-0.5mm]{}}

\newcommand{\onecm}{\mx[1cm]{}}

\newcommand{\bc}{\begin{center}}
\newcommand{\ec}{\end{center}}
\newcommand{\beq}{\begin{equation}}
\newcommand{\eeq}{\end{equation}}

\newcommand{\be}{\begin{enumerate}}
\newcommand{\ee}{\end{enumerate}}
\newcommand{\bi}{\begin{itemize}}
\newcommand{\ei}{\end{itemize}}
\newcommand{\bd}{\begin{description}}
\newcommand{\ed}{\end{description}}
\newcommand{\beqn}{\begin{equation}}
\newcommand{\eeqn}{\end{equation}}

\newcommand{\beqna}{\begin{eqnarray}}
\newcommand{\eeqna}{\end{eqnarray}}
\newcommand{\beqnas}{\begin{eqnarray*}}
\newcommand{\eeqnas}{\end{eqnarray*}}
\newcommand{\beqnaa}{$$\begin{array}{rcll}}  
\newcommand{\eeqnaa}{\end{array}$$}  
\newcommand{\beqnal}{$$\begin{array}{l}}  
\newcommand{\eeqnal}{\end{array}$$}  
\newcommand{\beqnana}{$$\begin{array}{lrcll}}  
\newcommand{\eeqnana}{\end{array}$$}  
\newcommand{\btbl}[1]{\begin{center}\begin{tabular}{#1}}
\newcommand{\etbl}{\end{tabular}\end{center}}
\newcommand{\beqnc}{$$\begin{array}{rclcl}}
\newcommand{\eeqnc}{\end{array}$$}
\newcommand{\fn}{\footnote}

\newcommand{\Section}[1]{{\section{{\pfont #1}}}}

\newcommand{\Subsection}[1]{\bigskip{\large \subsection{#1}}}

\newcommand{\prf}{{\sc Proof. }}

\newtheorem{dclprop}{{\sc Proposition}} 
\newtheorem{dclprops}{{\sc Proposition}}[subsection] 
\newtheorem{dclbigthm}[dclprop]{THEOREM}

\makeatletter
\def\thmlabel#1{\@bsphack\if@filesw {\let\thepage\relax
\xdef\@gtempa{\write\@auxout{\string
\newlabel{#1}{{\@Roman{\@currentlabel}}{\thepage}}}}}\@gtempa
\if@nobreak \ifvmode\nobreak\fi\fi\fi\@esphack}
\makeatother

\newtheorem{dclthm}[dclprop]{{\sc Theorem}}   
\newtheorem{dclthms}[dclprops]{{\sc Theorem}}   

\newtheorem{dcllem}[dclprop]{{\sc Lemma}}
\newtheorem{dcllems}[dclprops]{{\sc Lemma}} 
\newtheorem{dclsublem}[dclprop]{{\sc Sublemma}}
\newtheorem{dclcor}[dclprop]{{\sc Corollary}}
\newtheorem{dclcors}[dclprops]{{\sc Corollary}} 
\newtheorem{dcldfn}[dclprop]{{\sc Definition}}
\newtheorem{dcldfns}[dclprops]{{\sc Definition}}
\newtheorem{dclasss}[dclprops]{{\bf Assumption}}
\newtheorem{dclass}[dclprop]{{\bf Assumption}}

\newenvironment{prop}{\medskip\begin{dclprop}\sl}{\end{dclprop}}
\newenvironment{thm}{\medskip\begin{dclthm}\sl}{\end{dclthm}}

\newenvironment{cor}{\medskip\begin{dclcor}\sl}{\end{dclcor}}

\newenvironment{dfn}{\medskip\begin{dcldfn}\sl}{\end{dcldfn}}

\newenvironment{lem}{\medskip\begin{dcllem}\sl}{\end{dcllem}}

\newcommand{\bsl}{\begin{verse}\sl}
\newcommand{\esl}{\end{verse}}

\newtheorem{exxs}[dclprop]{Exercises}

\newenvironment{exercises-with-preamble}{\begin{exxs}\rm}{\end{exxs}}

\newcommand{\bthm}{\begin{thm}}
\newcommand{\ethm}{\end{thm}}
\newcommand{\bprop}{\begin{prop}}
\newcommand{\eprop}{\end{prop}}
\newcommand{\blem}{\begin{lem}}
\newcommand{\elem}{\end{lem}}
\newcommand{\bcor}{\begin{cor}}
\newcommand{\ecor}{\end{cor}}
\newcommand{\bdfn}{\begin{dfn}}
\newcommand{\edfn}{\end{dfn}}

\newcommand{\bz}{\begin{quote}\small}
\newcommand{\ez}{\end{quote}}

\newcommand{\einference}[2]  
  {\shortstack
      {$ #1 $\\ \mbox{}\\ $ #2 $}}

\newlength{\txtlth}
\newlength{\txtht}

\usepackage{amsmath,amssymb,oldlfont}
\usepackage{proof}
\usepackage{colordvi}
\usepackage{epsfig}
\newcommand{\st}{\hbox{{\small\bf ST}}}
\newcommand{\stv}{\hbox{{\small\bf STV}}}

\newcommand{\str}{\hbox{{\small\bf STR}}}

\begin{document}
\thispagestyle{empty}

\begin{center}
{\Large\pfont
A generic imperative language for polynomial time}\\[2mm]
Daniel Leivant {\tt (leivant@indiana.edu)}\\[1mm]
{\small SICE, Indiana University and IRIF, Universit\'{e} Paris-Diderot}\\[1cm]
\end{center}


\begin{abstract}
The ramification method in Implicit Computational Complexity has been
associated with functional programming, but adapting it to
generic imperative programming is highly desirable, given the wider
algorithmic applicability of imperative programming.

We introduce a new approach to ramification which, among other benefits,
adapts readily to fully general imperative programming.  
The novelty is in ramifying
finite second-order objects, namely finite structures,
rather than ramifying elements of free algebras. 

In so doing we bridge between Implicit Complexity's type theoretic 
characterizations of feasibility,
and the data-flow approach of Static Analysis.
\\[3mm]
\noindent
\zero\bxbf{Keywords:} Finite partial structures, 
structure transformation, imperative programs, 
ramification, polynomial time, implicit computational complexity.
\end{abstract}

\setcounter{page}{0}

\newpage

\Section{Introduction}

The analysis and certification of resource requirements of computer programs
is of obvious practical as well as foundational importance.
Of particular interest is the certification of {\em feasibility,} 
commonly identified with polynomial time (PTime),
i.e.\ algorithms that terminate in a number of steps polynomial 
in the size of the input.  

Two main approaches to the development of PTime-certified 
programs have been Implicit Computational Complexity (ICC)
and {\em Static Analysis} (SA).
ICC strives to characterize complexity 
by means that do not refer directly to resource usage.
A major strand of ICC has been the design of 
type systems that guarantee the feasibility of {\em declarative} programs,
based notably on data ramification and linear types.
In contrast SA attempts to
ascertain the feasibility of {\em imperative}
programs, primarily via data-flow analysis at compile time.
SA has the advantage of applying readily
to algorithms in which inert data play a direct role, 
such as nodes of a graph or raw data in memory management.
Consequently, bridging the two approaches, notably via type systems 
for imperative programs, is highly desirable. 

The main ICC approach to PTime originates 
with Cobham's characterization of PTime in terms
of bounded recurrence \cite{Cobham65}.
Advances in this area since the 1990's
were focused on mechanisms that limit data-duplication (linearity),
data-growth (non-size-increase), and nesting of iteration (predicativity)
(see \cite{Schimanski08} for a survey).
In its basic form, {\em predicative recurrence,} 
also known as {\em ramified recurrence,}
refers to computational ranks,
and requires that every iteration is paced by data-objects of
higher rank than the output produced.
This prevents the use of
non-trivial computed functions as the step-functions of recurrence.


\Subsection{The ramification method}
One main strand of implicit computational complexity (ICC) has been 
ramified recurrence,
also known as ``ranked'', ``stratified'', 
``predicative'', and ``normal/safe'' recurrence 
(i.e.\ primitive recursion over free algebras).
The approach has raised the hope for a practical delineation 
of feasible computing within
the primitive recursive functions, which arguably include all 
functions of interest.
This idea goes back to Ritchie and Cobham \cite{Ritchie63,Cobham65},
who introduced recurrence restricted explicitly by bounding conditions. 
although the characterizations they obtained 
use one form of bounded resources (i.e.\ output size) 
to delineate another form (i.e.\ time/space resources), 
they proved useful, for example in suggesting
complexity measures for higher-order functionals \cite{Constable73}.

A more foundational approach was initiated by 
a proof theoretic characterization of FPtime
based on a distinction between two
second order definitions of the natural numbers
\cite{Leivant94}.
This triggered\fn{Personal communication with Steve Bellantoni} 
the ``safe recurrence" characterization of PTime by
Bellantoni and Cook \cite{Bellantonic92},
as well as the formally more general approach of \cite{Leivant94}.

In fact, the ramification of programs
can be traced to the type theory of Fundamenta Mathematicae 
\cite{WhiteheadR12} whose simplest form is conveyed 
in sch\"{u}tte's ramified second order logic \cite{Schutte-proof}.  
The idea is to prevent impredicative set quantification.
A formula $F \equiv \forall s \; F_0[s]$ implies in second order 
logic\fn{We write
\zero $\lambda x \, G$ for the set consisting of those elements $a$ for which
$G$ is true under the valuation $x \mapsto a$.}
$F_0[\grl x.g]$, for any formula $G$, even if it is more complex than $F$.
Thus, the truth of $f$ depends on
the truth of $f_0[\grl x.g]$, which may itself have $f$ as a subformula.
While this form of circularity
is generally admitted as sound, it does raise
onthological and epistemological questions \cite{Kreisel60},
and implies a dramatic increase in definitional, 
computational, and proof-theoretic complexities of second-order
over first-order logic.
Sch\"{u}tte's ramified second-order logic blocks impredicative inferences
of the kind above by assigning a rank to each set definition, starting
with set variables. In particular, the rank of 
$f \equiv \forall s \;. f_0$, is larger than the rank of $s$,
so $s$ cannot be instantiated to $f$ (or any formula having $f$ as a 
subformula).

Schutte's ramification of sets yields a separate definition
of \dN\ for each rank $k$:\\
 	$(\forall s \text{ of rank } k) \; c[s] \ra s(x)$
where $c[s]$ is $s(\ttzero) \; \wedge \; 
 		\forall z . \; s(z) \sra s(\tts(z))$
analogously, ramified recurrence allows the definition
of a function $f: \; \dN \sra \dN$ with
output of rank $k$ only if the recurrence argument is of 
rank $ > k$.

Ramified recurrence has been used to obtain machine-independent 
characterizations of several major complexity classes, such as
polynomial time \cite{Bellantonic92,Leivant94} and
polynomial space \cite{LeivantM-ramifiedII,Oitavem08}, 
as well as
alternating log time\cite{Bloch92,Leivantm00},
alternating poly-log time \cite{Bloch92},
NC \cite{Leivant-nc,Oitavem04},
logarithmic space \cite{Oitavem10},
monotonic PTime \cite{DasO18},
linear space \cite{Leivant-popl93,Handley92,Leivant94},
NP \cite{Bellantoni92,Oitavem11},
the poly-time hierarchy \cite{Bellantoni94},
exponential time \cite{Clote97},
Kalmar-elementary resources \cite{Leivant-ramifiedIII}, 
and
probabilistic polynomial time \cite{LagoT15}.
The method is all the more of interest given
the roots of ramification in the foundations of mathematics
\cite{WhiteheadR12,Schutte-proof}, thus bridging
abstraction levels in set-theory and type-theory
to computational complexity classes.

While the ramification method 
has the markings of practicality as well as a foundational pedigree,
implementation, in particular for algorithms over finite structures,
has been problematic.
Notwithstanding its strengths,
the ramification method has, unfortunately, failed to date
to evolve into an effective practical method for static certification of
computational resources. 
One limitation is unavoidable:  {\em any} effective characterization of PTime
is necessarily {\em extensional:} even if every PTime {\em function}
is capture, not every PTime {\em algorithm} can be,
not even via effective enumeration:

\bthm\label{thm:ptime-notSD} Let $L$ be a Turing-complete programming language,
whose programs simulates Turing machines within PTime overhead.\fn{As do all 
programming languages in use.}
Let $L^p$ consist of the PTime $L$-programs.
Then $L^p$ is not semi-decidable.
\ethm
A proof is given in Appendix 1.\fn{The undecidability of PTime 
is folklore, but non-semi-decidability seems to be a new observation.}

But the trouble with ramification is that a number of fundamental algorithms 
elude it.
Caseiro observed \cite{Caseiro97} that important algorithms,
notably for sorting, do use recursively defined step functions,
and yet are in PTime, because those step functions are not increasing
the size of their principal argument. Hofmann built on that observation
\cite{Hofmann03,AehligBHS04}, and developed a type system for
non-size-increasing PTime functions, based on linearity and
an explicit account of information unit.
Unfortunately, the functions obtained
are all non-size-increasing, leaving open the meshing of
meshing these results with full PTime.

These difficulties seem related to foundational issues.
For one, confining ramification to recurrence ties it to
inductive data, thereby dissociating it from
finite data-structures. More generally, the focus on declarative programming
complicates direct access to memory which lies at the heart of
many feasible algorithms.  To overcome these limitations one need 
a germane applicability of the ramification method to imperative programming,
which is what we are proposing.

While ramification (and closely-related methods) has
been considered for imperative programming, they have focused on
restricted data-types, such as
stacks \cite{KristiansenN04,Kristiansen01},
words in \cite{Marion11,MarionP14},
and finite graphs in \cite{LeivantM13},
in constrained contexts that do not seem to generalize naturally.

The challenge is thus to design programming languages that accommodate
PTime algorithmic methods as comprehensively and flexibly as possible.
Given that PTime is often related to micro-code and memory management,
a PTime certification framework that applies to
imperative programming, and encompasses both inductive types
and micro-level data, should be particularly desirable.
We propose here just such a framework.

\Subsection{Static analysis}

We mentioned that leading approaches to resource certification
include Implicit Computational Complexity (ICC)
and Static Analysis (SA).
SA is algorithmic in nature: it focuses on a broad programming language 
of choice, and seeks to determine by syntactic means
whether given programs in that language are feasible.
This is in contrast to ICC, which attempts to create from the outset 
specialized programming languages or methods that delineate a complexity class. 
Thus, SA's focus is on compile time, making no demand on the programmer;
whereas ICC is a language-design discipline, that seeks to confine
programming to a safe regime.
The distinction between SA and ICC is not clear cut, however:
the syntactic restrictions embedded in a programming language
designed by ICC, might be derived by a smart compiler; 
conversely, program properties sought by an SA algorithm
might be broad enough to be rephrased as delineating a programming 
language.
An example of the SA approach is the line of research that refers to
the Meyer-Ritchie characterization of primitive recursion 
by imperative ``loop"-programs over \dN\ \cite{MeyerR67},
seeking algorithms for ascertaining the PTime termination of 
such programs \cite{Kristiansen01,KristiansenN04,
BenamramJK08,Benamram10,BenamramH19}.



\Subsection{A new approach to ramification}

One novelty of our approach is the use of
finite partial-functions as fundamental data-objects.
That choice leads to using data-consumption as a generic form 
of recurrence, capturing primitive recursive complexity
\cite{LeivantM19}.
However, a simple-minded ramification of data-depletion is fruitless,
because it blocks all forms of duplication, resulting in linear-time
programs.\fn{This issue does not come up with traditional ramified recurrence, 
due to the free repetition of variables in function composition.}
We resolve this snag by ramifying
all data, and trading off size-reduction of depleted data
with size-increase of non-depleted data within the same rank.

The rest of the paper is organized as follows.
Section 2 introduces the use of finite partial (fp) functions as basic data,
and describes an imperative programming language of primitive-recursive
complexity, based on fp-functions depletion 
\cite{Leivant19,LeivantM19}.
Section 3 introduces the ramified programming language \str,
shows that it is sound for PTime, and presents examples that illustrate
the methods and scope of the language.
Some of those examples are  used in \S 4 to prove that
\str\ is extensionally complete for PTime, i.e.\ has a program for every
PTime mapping between finite partial-structures.
The conclusion (\S 5) argues that the method is particularly amenable to
serve as a synthesis of an ICC core language, whose implementation
can be refined using SA methods.



\zero\\[4mm]

\Section{Programs for transformation of structured data}

\Subsection{Finite functions as data objects}

Basic data objects come in two forms: structure-less
``points", such as the nodes of graphs, versus elements of inductive
data, such as natural numbers and strings over an alphabet.  The former have
no independent computational content, whereas the computational
nature of the latter is conveyed by the recursive definition 
of the entire data type to which they belong, via the corresponding
recurrence operators. This dichotomy is antithetical, however,
to an ancient alternative approach that takes individual inductive 
data objects, such as natural numbers,
to be finite structures on their own, whose computational
behavior is governed by their internal makings 
\cite{EuclidElements,Mayberry11}.
Under this approach, computing over
inductive data is reduced to operating over finite structures,
and functions over inductive data are construed as mappings
between finite structures.
 
Embracing this approach yields a number of benefits. 
First, we obtain a common ``hardware-based" platform for programming 
not only within finite structures, but also for the transformation 
of inductive data.
Conjoining these two provides a common platform
for microcode and high-level programming constructs, 
In particular, the depletion of
natural numbers the drives the schema of recurrence over \dN, is
generalized here to the depletion of finite functions as loop variants.

Yet another benefit of our approach is the generalization of
the ramification method in implicit computational complexity
to imperative programs over finite structures.
The step of ramifying recurrence over \dN\ (or $\{0,1\}^*$) to
obtain a PTime form of recurrence, is reformulated here
in an imperative context which provides greater algorithmic
flexibility, and deals with algorithms that are difficult
to express using traditional ramified recurrence.




Focusing on finite structures may seem akin to
finite model theory, with finite structures taken to be
particular Tarskian structures.  But once we construe finite structures 
as data objects, we obtain an infinite data type, such as \dN, 
to be a collection of particular finite structures, and computing over \dN\ as
a process of transforming those structures.
For instance, a program for string reversal
takes as input a string-as-finite-structure
and yields another string-as-finite-structure.

\Subsection{Finite partial-structures}

We take our basic data-objects to be {\em finite partial-functions}
(fp-functions) in the following sense.
We posit a denumerable set $A$ of {\em atoms,}
i.e.\ unspecified and unstructured objects.
To accommodate in due course non-denoting terms
we extend $A$ to a set $A_\bot \df A \cup \{\bot\}$,
where $\bot$ is a fresh object intended to denote ``undefined."  
The elements of $A$ are the {\em standard} elements of $A_\bot$.
A {\em ($k$-ary) fp-function} is a function $F:\; A_\bot^k \ra A_\bot$
which satisfies:
\be
\li $F(\vec{a}) = \bot$ for all $\vec{a} \in A^k$ except
for a $\vec{a}$ in a finite set $A_F$, dubbed here the {\em domain} of $F$.
\li $F$ is strict: $F(\vec{a}) = \bot$ whenever $\bot$ is one of the
arguments $\vec{a}$.
\ee
An {\em entry} of $F$ is a tuple $\lng a_1 \ldots a_k ,b \rng$ 
where $b = F(a_1,\ldots, a_k) \neq \bot$.
The {\em image} of $F$ is the set $\{ b \in A \mid b = F(\vec{a}) 
\text{ for some } \vec{a} \in A^k\,\}$.

Function partiality provides a natural representation of finite relations
over $A$ by partial functions, avoiding {\em ad hoc} constants.
Namely, a finite $k$-ary relation $R$ over $A$ ($k >0$)
is represented by the fp-functions
$$
\grx_R^k(a_1, \ldots, a_k) \; = \; \text{ if } 
	R(a_1, \ldots, a_k) \text{ then } a_1 \text{ else $\bot$ }
$$
Conversely, any partial $k$-ary function $F$ over $A$ determines the
$k$-ary relation
$$
R_F \; = \; \{\lng \vec{a} \rng \in A^k \mid \; F(\vec{a}) \;
        \text{is defined} \; \}
$$
A {\em vocabulary} is a finite set $V$
of function-identifiers, referred to
as $V$-ids, where each $\bff\!\in\! V$ is assigned an {\em arity}
$\ar(\bff) \geq 0$. We optionally right-superscript an identifier by its arity,
when convenient.
We refer to nullary $V$-ids as {\em tokens}
and to identifiers of positive arities as {\em pointers.} 
Our default is to use type-writer symbols for identifiers:
$\tta,\ttb,\tte, ...\;$ for tokens and $\;\tts,\ttf,\ttg,\ttzero,\ttone ...\;$ 
for pointers.
The distinction between tokens and pointers is computationally
significant, because  (non-nullary) functions can serve as unbounded memory,
atoms cannot.  For a vocabulary $V$, we write $V^0$ for the
set of tokens, and $V^+$ for the set of pointers.

An {\em fp-structure over $V$}, or briefly a {\em $V$-structure},
is a mapping \grs\ that
to each ${\bff\,^k} \in V$,
assigns a $k$-ary fp-function $\grs(\bff)$,
said to be a {\em component} of \grs.
The intention is to identify isomorphic fp-structures,
but That intention may be left implicit without complicating matters 
with perpetual references to equivalence classes.
Note that a tuple $\vec{\grs}= (\grs_1, \ldots, \grs_k )$ of fp-structures is
representable as a single structure, defined as
the union $\sbcup_{1 \leq i \leq k}\, \grs_i$  over
the disjoint union of the vocabularies $V_i$.


The {\em domain} (respectively, {\em range}) of an fp-structure \grs\ is the union of the 
domains (ranges) of its components,
and its {\em scope} is the union of its domain and its range. 
If \grs\ is a $V$-structure, and \grt\ a $W$-structure, $W \supseteq V$,
then \grt\ is an {\em expansion} of \grs\ (to $W$), and \grs\ a {\em reduct}
of $\grt$ (to $V$), if the two structures have
identical interpretations for each identifier in $V$.
%

Given $\bff \in V$ and a $V$-structure \grs, 
the {\em size of \bff\ in \grs,} denoted $|\bff\,|_\sigma$, 
is the number of entries in $\grs(\bff)$.
For $Q \subseteq V$ the {\em size of $Q$ in \grs,} denoted $|Q|_\sigma$,
is $\sum \{\, |\bff\,|_\sigma \; \mid \bff\in Q \}$.
We refer to $|V|_\sigma$ as the {\em size of \grs.}


\Subsection{Terms}\label{subsec:terms}

Given a vocabulary $V$, the set $\bxbf{Tm}_V$ of {\em $V$-terms} is 
generated inductively, as usual:
$\bgrw \in \bxbf{Tm}_V$;
and if $\bff\,^k\in V$ ($k \geq 0$),
and $\bft_1, \ldots , \bft_k \in \bxbf{Tm}_V$ then
$\bff\bft_1 \cdots \bft_k \, \in \bxbf{Tm}_V$.
A term \bft\ is {\em standard} if $\bfgrw$ does not occur in it.
Note that tokens assume here the traditional role of
program variables. In other words, we do not distinguish between
an underlying structure and a store.

We write function application in
formal terms without parentheses and commas, as in $\bff xy$ or $\bff \vec{x}$.
Also, we implicitly posit that the arity of a function matches the number of
arguments displayed; thus writing $f^k\vec{a}$ assumes that
$\vec{a}$ is a vector of length $k$,
and $f\vec{a}$ (with no superscript) that 
the vector $\vec{a}$ is as long as $f$'s arity.




Given a $V$-structure \grs\
the {\em value} of a $V$-term \bft\ in \grs,
denoted $\grs(\bft)$, is obtained by recurrence on \bft: 
$\grs(\bfgrw) = \bot$; and for $\bff\,^k \in V$,
$\grs(\bff \bft_1 \cdots \bft_k) =
	\grs(\bff)(\grs(\bft_1), \ldots , \;
			\grs(\bft_k))$.
%
%
An atom $a \in A$ is {\em $V$-accessible in \grs}
if it is the value in \grs\ of some $V$-term.
A $V$-structure \grs\ is {\em accessible} if every atom in the 
domain of \grs\ is $V$-accessible
(and therefore every atom in the range of \grs\ is also accessible).


If every atom in the range of an accessible structure \grs\ is the value
of a {\em unique} $V$-term then \grs\ is {\em free}.
It is not hard to see that
an accessible $V$-structure \grs\ is free 
iff there is a finite set $T$ of $V$-terms, closed under
taking sub-terms, such that the valuation function $\grs:\; T \ra A$
is injective.

%

%


If \bfq\ is a standard $V$-term, and $T$ consists of the
sub-terms of \bfq, then we write $\grf_{\scriptbfq}$ for the resulting
{\em free} fp-structure, with a token $\blacksquare$
designating the term as a whole. Examples are in Appendix 2.

%

%

\Subsection{Structure updates}
Fix a vocabulary $V$.
We consider the following three basic operations on $V$-structures. 
In each case we indicate how an input $W$-structure \grs\ (with $W \supseteq V$)
is transformed by the operation into a $W$-structure $\grs'$ 
that differs from \grs\ only as indicated.
\bd
\item{1.} A {\em $V$-extension} is a phrase
$\bff\, \bft_1 \cdots \bft_k \sdownarrow \bfq$ where
\ttq\ and each $\bft_i$ are standard $V$-terms.
The intent is that if
$\grs(\bff\, \bft_1 \cdots \bft_k) = \bot$, then
$\grs'(\bff\, \bft_1 \cdots \bft_k) = \grs(\bfq)$.
Thus, $\grs'$ is identical to \grs\ if
$\grs(\bff\, \bft_1 \cdots \bft_k)$ is defined.

\item{2.} A {\em $V$-contraction,} the dual of an extension,
is a phrase of the form
$\bff\bft_1 \cdots \bft_k \suparrow.$
The intent is that
$\grs'(\bff)(\grs(\bft_1), \ldots, \grs(\bft_k)) = \bot$.
Note that this removes the entry\\
$ \lng \grs(\bft_1) ,\ldots ,  \grs(\bft_k),
\grs(\bff\bft_1 \cdots \bft_k) \rng $ (if defined) from $\grs(\bff)$,
but not from $\grs(\bfg)$ for other identifiers \bfg.

\item{3.} A {\em $V$-inception} is a phrase of the form 
$\bfc \ssDownarrow$, where \bfc\ is a $V$-token.
A common alternative notation is $\bfc := \bxbf{new}$. 
The intent is that $\grs'$ is identical to \grs,
except that if $\grs(\bfc) = \bot$, then $\grs'(\bfc)$ is
an atom not in the scope of \grs.

\item{\mywhite{4.}}
In all cases we omit the reference to the vocabulary $V$ if
in no danger of confusion.
\ed

We refer to extensions and contractions as {\em revisions,}
and to revisions and inceptions as {\em updates.}
The identifiers \bff\ in the revision templates above are the
revision's {\em eigen-id.} 
An extension [contraction] is {\em active (in \grs)}
if, when triggered in \grs, it adds [respectively, removes] an entry from its eigen-id.

\noindent
\bxbf{Remarks.}
\be

\li
An assignment $\bff\vec{\bft} := \bfq$ can be programmed by 
composing extensions and contractions:
\begin{equation}\label{eq:assign}
\bfb \sdownarrow \bfq; \;\;
\bff\vec{\bft}\uparrow; \;\;
\bff\vec{\bft} \sdownarrow \bfb; \;\;
\bfb\suparrow
\end{equation}
where \bfb\ is a fresh token which memorizes the atom denoted by \bfq,
in case the contraction renders it inaccessible.

\li
Inception does not have a dual operation, since atoms can be released from
a structure by repeated contractions.
\li
A more general form of inception, with a fresh atom assigned
to an arbitrary term $\bft$,
may be defined by
$$ \bfb \sDownarrow; \;\; \bft \sdownarrow \bfb; 
		\;\; \bfb \suparrow	\qquad \qquad
		\text{(\bfb\ a fresh token)}
$$
\ee

\Subsection{Programs for transformation of fp-structures}\label{subsec:stv}

Fix a vocabulary $V$.
A {\em $V$-guard} is a boolean combination of $V$-equations.
A {\em $V$-variant} is a set of $V$ pointers.
The imperative programming language $\stv$, consists of the
{\em $V$-programs} inductively generated as follows \cite{Leivant19,LeivantM19},
(We omit the references to $V$ if in no danger of confusion.)


\bd
\li[\bxbf{Update}] A $V$-update is a $V$-program.
\ed
If $P$ and $Q$ are programs, then so are  the following.
\bd
\li[\bxbf{Composition}] \quad $\;P;\,Q\;$ 
\li[\bxbf{Branching}]\qquad  
	$\;\bxbf{if}\,[G]\,\{P\}\,\{Q\}\;$ \quad ($G$ a $V$-guard)
\li[\bxbf{Iteration}] \qquad $\; \bxbf{do}\;[G]\,[T]\; \{P\}\;$ 
		\quad ($G$ a $V$-guard, $T$ a $V$-variant)
\ed


The denotational semantics of the Iteration template above
%
%
%
calls for the loop's body $P$ to be entered initially
if $G$  is true in the initial structure \grs, and re-enter if
\be
\li $G$ is true for the current structure, and
\li The size of the variant $T$ is reduced, that is: 
the execution of the latest pass through $P$ executes more 
active contractions than active extensions of the variant.
\ee
In particular, the loop is existed if the variant $T$ is depleted.
A formal definition of this semantics in terms of configurations 
and execution traces is routine.

From the vantage point of language design,
termination by depletion is a common practice.
However, keeping track of the balance of active extensions and active 
contractions requires an unbounded counter.  
If this, for some reason, is to be avoided, 
one can resort to more local forms of control. Here are two
such options.
\be
\li 
Syntactically, require that loops $\bxbf{do}[G][T]\{P\}$ have no 
extension of $T$ in $P$. 
Semantically, scale down the depletion condition of \stv\ to 
just one active contraction. 
This implementation eliminates the need for unbounded counters
in an implementation of \stv\ to just one flag per loop.
The resulting variation of \stv\ still yields full primitive recursion 
\cite{LeivantM19}.
\li 
Define a {\em pod} to be the composition of updates (possibly
a single update).
Programs are then generated from pods as basic building blocks.

The semantics of iteration is defined in terms of pods, as follows.
Say that the execution of a pod-occurrence is {\em positive} [respectively, 
{\em negative}] in \grs\ 
the number of active extensions is larger [respectively,
smaller] than the number of active contractions.

The semantics of $\bxbf{do}[G][T]\{P\}$ is calls, then, to exit the loop
if the latest pass has no positive execution of any pod, and has
at least one negative one.
This reduces the unbounded counter of \stv\ to local counters for each pod.
\ee


The {\em resources} of a $V$-program $p$ are defined in terms of
the size of an input fp-structure, i.e.\
the total number of entries (not of atoms).
For a function $f: \; \dN \sra \dN$ we say that
program $p$ is in $\bxrm{space}(f)$ if there
is a constant $c>0$ such that for all fp-structures \grs\
the size of structures $\grs'$ in the execution trace of $P$
for input \grs\ is $\leq c \cdot f (|\grs|)$.
$P$ is in PSpace if it is in $\bxrm{space}(\grl n.n^k)$ for some $k$.
$P$ is in $\bxrm{time}(f)$ if there
is a $c$ such that for all fp-structures \grs\
as input, $P$ terminates using an execution trace of length 
$\leq c \cdot f (|\grs|)$.  
$P$ is in PTime if it runs in time $o(\grl n.n^k)$ for some $k$.
We focus here on programs as transducers.
a partial mapping  $\grf:\; \gothc \pa \gothc'$ from a class
\gothc\ of $V$-structures to a class $\gothc'$ of $V'$-structures
is {\em computed} by a program $P$, over vocabulary $W\supseteq V\cup V'$,
if for every
$\grs \in \gothc$, $\grs^W \ra_P \grs'$ for some $W$-expansion $\grs'$
of $\grf(\grs)$, where $\grs^W$ is the trivial expansion of \grs\
to $W$, with every $\bff \in w\sminus v$ interpreted as empty 
(i.e.\ undefined for all arguments).



\bthm\label{thm:stv=pr-resources}\zero\bxrm{\cite{LeivantM19}}
Every \bxbf{\stv}-program runs in time and space primitive-recursive in the
size of the input. 
\ethm

\Section{Feasible termination}

\Subsection{Programs for generic PTime}

We define the programming language \str, which modifies \stv\ 
by the ramification of variants.
We depart from traditional ramification, which assigns ranks to 
first-order inductive data-objects, and ramify instead the
loop variants, i.e.\ second-order objects.
This is in direct agreement
with the ramification of second-order logic \cite{Schutte-proof} and,
more broadly, of type theory \cite{WhiteheadR12,Palmgren18}.

Ramification is a classification of the computational
powers of objects that drive iteration.
we use natural numbers for ranks.

A {\em ramified vocabulary} is a pair $(V,\grr)$ where $V$ is
a vocabulary and $\grr: \; V \ra \dN$. we refer to $\grr(\bff)$ 
as the {\em rank} of \bff, and let $V_r \df \{ \bff \in v \mid \grr(\bff) = r$.
A {\em variant (of rank $r$)}, for a ramified vocabulary $(V,\succeq)$,
is a set $T \subseteq V_r$.

The syntax of \str\ is identical to \stv, except for the iteration clause,
which is replaced by 

\bi
\li[\bxbf{Ramified Iteration}]
If $G$ is a guard, $T$ a variant of rank $r$,
$\bxbf{do}[G][T]\{P\}$ is a program.
\ei

The semantics of this iterative program has the 
loop's body $P$ re-entered when in configuration (i.e.\ fp-structure) \grs\ 
if the two conditions of \stv\ above are supplemented by a third:
\be
\li $G$ is true in \grs.
\li $|T|$ is shrunk in the previous pass, 
i.e. the number of active contractions
of components of $T$ in $P$ exceeds the number of active extensions.
\li For $j \geq r$, $|V_j|$ does not grow; that
is, the total number of active extensions of pointers of rank $j$ does not
exceed the total number of active contraction.
\ee

\noindent
\bxbf{Remarks.}
\be
\li A loop with a variant $T$ of rank $r$
will be re-entered after decreasing $T$ even if
that decrease is offset by {\em extensions} of 
pointers in $V_r - T$.  Allowing such extensions is essential: 
programs in which loops of rank $r$ cannot extend $V_r-T$ 
execute in linear time,
as can easily be seen by structural induction.

\li We caution the reader familiar with existing approaches to 
ramified recurrence that 
our ranks are properties of function-identifiers,
and not of atoms, fp-functions, or terms. Moreover, no ranking
for atoms or functions is
inherited from the ranking of function-ids: an fp-function
may be the value of distinct identifiers, possibly of different ranks.
In particular, there is no rank-driven restriction on inceptions or
extensions: the function-entries created have no rank,
e.g.\ an extension $\bff \bfc \sdownarrow \bfq$ may have \bff\ 
of rank 0 whereas \bfq\ refers to arbitrarily large ranks.

\li The condition on non-increase of $V_j$ for $j > r$ has no
parallel in ramified recurrence, but is needed for imperative
programs, in which every variable may be considered an output-variable.

\li If unbounded counters for the size of ranks are to be avoided,
they can be replaced by local counters for pods, as in \S\ref{subsec:stv}.
The simpler approach described there, of disallowing extensions altogether,
is not available for \str,
because (as observed) extensions of an iteration's rank is essential to permit
data-transfers within that rank.
\ee

\Subsection{PTime soundness of \str}

\bthm\label{lem:f-programs-sound}
For each \str-program $P$ with loop ranks $\leq \ell$,
there is a positive\fn{I.e. defined without subtraction} 
polynomial $M_P[n_0\ldots n_\ell]$ such that for all $V$-structures \grs\  
$$
\bxrm{Time}_P(\grs) \leq M_P[|\grs|_0\ldots |\grs|_\ell]
$$
\li Moreover, for each $j \leq \ell$ there is a positive polynomial
$Z_{P,j}[n_{j+1} \ldots n_\ell]$ such that
$$
\bxrm{Space}_{P,j}(\grs)
	\leq |\grs|_j \; 
		+ \;  Z_{P,j}[|\grs|_{j+1}, \ldots, |\grs|_\ell]
$$
\ethm
\prf
Parts 1 and 2 are proved by a simultaneous induction on $P$.
Non-trivial case: $P \; \equiv \; \bxbf{do}[G][T]\{Q\}$,
where (by the definition of programs) $T \subseteq V_r$ is decreasing in $Q$,
and $V_r,V_{r+1}, \ldots , V_\ell$ are each non-increasing in $Q$.
Suppose $\grs \rA_P \grt$, where
$$
\grs = \grs_0 \rA_Q \grs_1 \cdot\cdot\cdot \rA_Q \grs_k = \grt
$$
Since $T$ is decreasing in $Q$, we have
$k \leq |T|_\sigma \leq |\grs|_r$. For each $j\geq r$ 
$V_j$ is non-increasing in $P$,
so we take $Z_{P,j} \equiv 0$.

For $j= r-d$, $d= 0, \ldots, r$,  we proceed by a secondary induction on $d$.
The induction base $d=0$, i.e.\ $j=r$, is already proven.  
For the step, we have
$$\begin{array}{l}
\bxrm{Space}_{P,r-(d+1)}(\grs)\\
\zero \qquad \begin{array}{lll}
	= &  
		\max_{i<k} \bxrm{Space}_{Q,r-(d+1)}(\grs_i)\\ 
	\leq &  
		\max_{i<k} 
		Z_{Q,r-(d+1)}[|\grs_i|_{r-d}, \ldots, |\grs_i|_\ell]
			& \text{(by main IH)} \\
	\leq &  
		Z_{Q,r-(d+1)}[\bxrm{Space}_{P,r-d}(\grs),\;
			\ldots, \; \bxrm{Space}_{P,\ell}(\grs)]
			& \text{(by definition of $\bxrm{Space}_{P,j}$}\\
	&& \text{and since each $Z_{Q,j}$ is positive)}\\
	\leq &  
		Z_{Q,r-(d+1)}[A_{r-d},\, \ldots\, ,\, A_\ell]
			& \text{(by secondary IH)}\\
	\end{array}\\[3mm]
\zero\qquad\qquad \text{where} \quad  A_j \; \text{stands for} \;
		|\grs|_j +\,  Z_{P,j}[|\grs|_{j+1}, \ldots , |\grs|_\ell]   
\end{array}$$
So it suffices to take
$$
Z_{P,r-(d+1)}[n_{r-d}, \ldots , n_\ell]
	\; \dfr \;
	Z_{Q,r-(d+1)}[B_{r-d} \ldots B_{\ell}]
$$
where $B_j$ stands for  $n_j + Z_{P,j}[n_{j+1}, \ldots , n_\ell]$.


This concludes the inductive step for (2).

For the inductive step for (1), we have
\beqnaa
\bxrm{Time}_P(\grs) 
	& \leq & k 
		+ \sum_{i < k} \bxrm{Time}_Q(\grs_i)\\
	& \leq & k
		+ \sum_{i < k} M_Q(|\grs_i|_0,\, ... , \, |\grs_i|_\ell)
			& \text{by IH}\\
	& \leq & k 
		+ \sum_{i < k} M_Q[A_0,\, ... , \, A_\ell]
			& \text{with the $A_j$'s above} \\
	& \leq & |\grs|_r \; ( 1 +
		M_Q[A_0,\, ... , \, A_\ell])
\eeqnaa
So it suffices to take
$$
M_P(n) \; \dfr \;\; n_r (1 + M_Q[B_0, \ldots , B_\ell])  
$$
where the $B_i$'s are as above. 
\qed

From Lemma \ref{lem:f-programs-sound} we conclude:

\bthm\label{thm:f-sound}
Every program of \str\ runs in time polynomial in the size of the 
input structure.
\ethm 

\zero\\[-16mm]

\Subsection{Examples of \str-programs}


Aside of illustrating our ramification mechanism, 
the following examples will 
establish structural expansions (\S\S 5.1-5.3, to be used in \S\S 
\ref{subsec:closure} and \ref{subsec:stv-compl}),
consider arithmetic operations (\S 5.4, used in\S \ref{subsec:stv-compl}),
and code several sorting algorithms (\S 5.5) which are problematic
under the traditional ramification regime. 

\subsubsection{String duplication}\label{subsec:duplicate}

The following program has as input a token \tte\ and
unary pointers $\ttf_0$ and $\ttf_1$
of rank 1.  The intended output consists
of \tte\ and the unary pointers $\ttg_0, \ttg_1, \ttg'_0$ and $\ttg'_1$ 
of rank 0.
Termination is triggered solely by depletion of the
variant $\{\ttf_0,\ttf_1\}$,
whence an empty guard (i.e.\ \bxbf{true}).\fn{Termination by 
depletion is indeed frequent in imperative programming!}
Note that the loop's body executes a contraction of the variant,
unless the variant is empty.

$$\begin{array}{ll}
\zero\qquad \tta:= \tte;\\[1mm]
\zero\qquad \bxbf{do}\;[\;\;]\, [\ttf_0,\ttf_1]\\[1mm]
\zero\qquad\qquad        \bxbf{if}\; [\, !\ttf_0\tta\, ]\\[1mm]
\zero\qquad\qquad \;\;    \{ \, \lng \, \ttg_0\tta \downarrow \ttf_0\tta,
		\;\; \ttg'_0\tta \downarrow \ttf_0\tta,
		\;\; \tta \downarrow \ttf_0\tta,
		\;\; \ttf_0\tta \uparrow \, \rng \, \}\\[1mm]
\zero\qquad\qquad \;\;    \{ \, \lng \, \ttg_1\tta \downarrow \ttf_1\tta,
		\;\; \ttg'_1\tta \downarrow \ttf_1\tta,
		\;\; \tta \downarrow \ttf_1\tta,
		\;\; \ttf_1\tta \uparrow \, \rng \, \}\\[1mm]
\zero\qquad\qquad 
		\}
\end{array}$$

The program consumes the variant while creating two copies at a lower
rank,
but in fact a rank-1 copy $(\ttf'_0,\ttf'_1)$ of the variant
may be constructed as well:
$$
\zero\qquad \, \lng \, \ttg_0\tta \downarrow \ttf_0\tta,
		\;\; \ttg'_0\tta \downarrow \ttf_0\tta,
		\;\; \ttf'_0\tta \downarrow \ttf_0\tta,
		\;\; \tta \downarrow \ttf_0\tta,
		\;\; \ttf_0\tta \uparrow \, \rng 
$$
and similarly for $\ttf_1$. The variant still decreases with each pass,
while $V_1$ is non-increasing.  Of course, 
no more than a single rank-1 copy can be created, lest $V_1$ would increase.

The copy $\ttf'$ created must be syntactically different from the 
variant $\ttf$, but the loop above can be followed by a loop that
similarly renames $\ttf'$ to \ttf.  We shall refer to this sort
of variant re-creation as {\em spawning.}
In particular, given a chain $L = (a,e)$ and a guard $G$,
spawning in rank 0 allows a scan of $L$ for an atom that satisfies $G$,
while consuming $L$ as a variant and recreating it at the same time.

\subsubsection{Enumerators}\label{subsec:enumerator}

We refer to an implementation of lists that we dub {\em chain,}
consisting of a token \tte\ and a unary injective pointer \ttf.
The intent is to represent a list of
atoms as the denotations of
$\tta, \ttf\tta,\ttf\ttf\tta \ldots \ttf^{[i]}\tta \ldots$,
where $\grs(\ttf)$ is injective.
Since \grs\ is an fp-structure and $\grs(\ttf)$ injective, 
the chain must be finite.

A chain $(e,f)$ is an {\em enumerator} for an fp-structure \grs\ if for some $n$
$$
e,\,  f(e),\, f(f(e)),\,  \ldots,\, f^{[n]}(e)
$$
is a listing (possibly with repetitions)
of the {\em accessible} atoms of \grs, and
$f^{[n+1]}(e) = \bot$.

Let $(V,\grr)$ be a ramified vocabulary. We may assume that
$\grr$ uses only ranks $> 1$, since raising all ranks by 2
results in a ranking function equivalent to $\grr$ (i.e.\ yielding
the same domination relation on $V$).

We outline an \str-program $E_V$
that for a $V$-structure \grs\ as input yields an
expansion $\grs_E$ of \grs\ with
an enumerator $(\tta,\ttl)$ in rank 0 for \grs.

$E$ initializes $\ttl$ to a listing of $\grs(\tta)$ for $V$'s tokens \tta.

Let $m = \sum_{\scriptbff \in V} \; \ar(\bff)$.
$E$ iterates then its main cycle $C$, which collects accessible elements
that are not yet listed in \ttl, into $m$ copies of a unary cache \ttp\
of rank 0, with for each $\bff\in V$ of arity $k$ a block 
$B(\bff)$ of $k$ copies of \ttp\ dedicated to \bff.
Using the entire vocabulary $V$ as variant, 
$C$ takes each $\bff^k \in V$ in turn,
cycles through all $k$-tuples in $B(\bff)i$
and for each tuple appends
$\grs(\bff)\vec{a}$ to all $m$ copies of \ttp\ if it is not already in \ttp.
That cycling through $B(\bff)$ takes $B(\bff)$ as variant in rank 1,
using spawning to preserve $B(\bff)$ as needed.
When this process is completed for all $\bff \in V$, $C$
concatenates (any one of the copies of) \ttp\ to \ttl, 

The loop is exited by variant-depletion, when the cache \ttp\ 
remains empty at the end of $C$ (no new atom found).
If input \grs\ is free, then the enumerator \ttl\ is {\em monotone:}
for each term $\bfq = \bff^k\bft_1\cdots \bft_k$ the enumerator lists $\bft_i$ 
before \bfq.
\qed

\subsubsection{Arithmetic functions}\label{subsec:arith}

Natural numbers are taken to be the free structures $\grf_{\scriptbfn}$
for the unary numerals $\bfn \in \dN$, for a vocabulary with one
token (the ``zero") and on unary pointer (the ``successor"). 

Addition can be computed in rank 0, which should not be surprising since
it does not increase the (combined) size of the inputs.
Splicing one input onto the other is not quite acceptable syntactically,
since the two inputs are given with different successor identifiers.
But the sum of natural numbers $(\Box,\ttz,\tts)$ and
$(\Box_0,\ttz_0,\tts_0)$, both
of rank 0, can  be computed by a loop that uses $\tts$ as variant,
and appends the first input to the first, starting from $\Box$.

Note though that the first input is depleted in the process, and
that spawning (in the sense of \ref{subsec:duplicate}) is disallowed
since the first input is in rank 0.
Positing that the first input is in rank 1 enables spawning, whence
a re-use of the first input.

That is precisely what we need for a program for multiplication.
We take both inputs to be in rank 1. The second input is a variant
for an outer loop, that sets the output to $\grf_0$, and
then iterates an inner loop driven by the first 
input as variant, that adds itself to the input while spawning itself as well.

It is worthwhile to observe how ramification
blocks exponentiation, as predicted by Theorem \ref{thm:f-sound}.  
A simple program for exponentiation
iterates the doubling operation starting with 1.
We have seen that any $\grf_n$ in rank $t$ can be duplicated 
into any number of copies, but at most one of these can be in rank $t$,
and all others in ranks $<t$.  

As for an iteration of multiplication,
our program above for the product function takes two inputs in rank $t$,
yielding an output in rank $< t$, a process that can be repeated only
$t$ many times for any fixed rank $t$.

\subsubsection{Insertion Sort}

{\em Insertion sort} is a non-size-increasing algorithm, and consequently has
an un-ramified (i.e.\ single ranked)  \str-program.
In general, we construe sorting algorithms as taking a chain $L = (a,e)$
and a partial order relation $\leq$, and return a
chain $K = (b,f)$ listing the same atoms as $L$ without repetition,
and consistent with $\leq$, i.e.\ 
\beqnaa
e^{[n]}(a) \neq \grw & \text{implies} & e^{[n]}(a) = f^{[m]}(b)
	& \text{for some $m$}\\
f^{[m]}(a) \neq \grw & \text{implies} &  f^{[m]}(b) = e^{[n]}(a)
	& \text{for some $n$}\\
&& f^{[m]}(a) < f^{[m+1](a)} & \text{ for all $m$} 
\eeqnaa

Our program for Insertion Sort is:

$$\begin{array}{ll}
\ttb \sdownarrow \tta;\\[1mm]
\bxbf{do}\;[\;\;]\, [\tte] 
	\qquad \text{(\ttf\ depleted via a spawned copy)}\\[1mm]
\zero\qquad\qquad 
\{ \; \lng \;
	\ttf\tta \sdownarrow \ttf\ttc,\; 
	\ttf\ttc \suparrow,\;
	\ttf\ttc \sdownarrow \tta, \;
	\ttc \suparrow, \;
	\ttd \sdownarrow \tte\tta,\;
	\tte\tta \suparrow, \;
	\tta \suparrow,\;
	\tta \sdownarrow \ttd, \;
	\ttd \suparrow\; \rng
		\; \}
\end{array}$$

\xfig{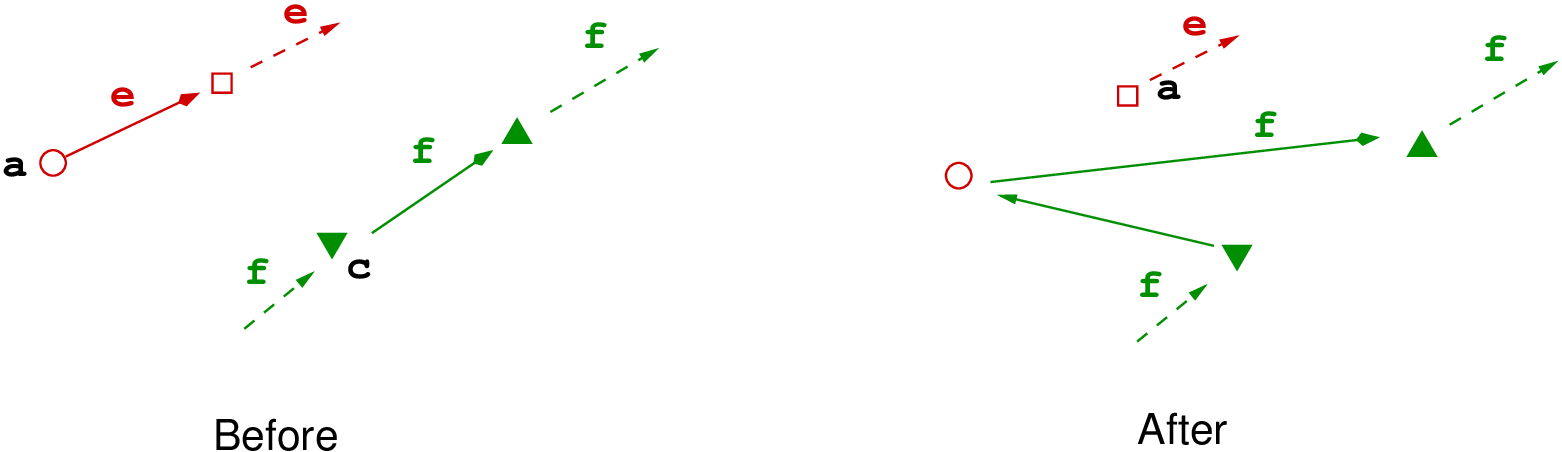}{32mm}

Note that the pod executes two extensions and two contractions,
while executing just one contraction on \tte.

\Section{Completeness of {\large\bf STR} for PTime}

\Subsection{Closure of \str\ under composition}\label{subsec:closure}

\bthm\label{thm:composition}
If partial-mappings $\grF_1,\grF_2$ between fp-structures are 
defined by \str-programs, the so is their composition
$\grF_1\circ\grF_2$.
\ethm
\prf
Given a transducer-program $P$ 
that uses ranks $r_1, \ldots , r_k$ for the input vocabulary,
we can modify $P$ to a program $P'$ that takes inputs that are all of 
a rank $r \geq r_1, \ldots, r_k$, copy the input into ranks 
$r_1, \ldots , r_k$, and then invokes $P$.
Dually, if the outputs of $P$ use ranks $q_1, \ldots, q_m$,
we can modify $P'$ to $P''$ that invokes $P'$ and then copies the outputs
into a rank $q_0 \leq q_1, \ldots , q_m$.

Let transducer-programs $P_1,P_2$ of \str\ 
compute $\grF_1,\grF_2$, respectively.
Suppose that the outputs of $P_1$ are the inputs of $P_2$
(so that composition may be defined).
As observed above,
we may assume that $P_i$'s inputs have a common rank
$t_i$, and the outputs have a common rank $s_i$ ($i=1,2$).
We wish to obtain an \stv\ transducer program for $\grF_1 \circ \grF_2$.

If $s_1 = t_2 +d$ where $d \geq 0$, let $P_2'$ be $P_2$ with all
ranks incremented by $d$. $P_2'$ is trivially a correct program of \stv,
with input of rank equal to the output rank of $P_1$.  So $P1;P_2'$
is a correct \stv-program for $\grF_1 \circ \grF_2$.

Otherwise, $t_2 = s_1 +d$, where $d > 0$. Let $P_1'$ be $P_1$ with
all ranks incremented by $d$. Then $P_1';P_2$ is a correct \stv-program
for $\grF_1 \circ \grF_2$.
\qed

\Subsection{Extensional completeness of \str\ for PTime}\label{subsec:stv-compl}

As noted in Theorem \ref{thm:ptime-notSD},
no programming language can be sound and complete for PTime {\em algorithms}.
\str\ is, however, {\em extensionally} complete for PTime.
This statement is best interpreted in relation to the programming language
\st\ of \cite{Leivant19}.  \st\ is simply \stv\ without the variants,
and it is easily seen to be Turing complete.


\bthm\label{thm:complete}
Every \st-program $P$ running in PTime is extensionally equivalent to some
\str\ program $P^*$; i.e.\ $P^*$ computes the same mapping between fp-structures
as $P$.
\ethm
\prf
Let $P$ be an \st-program over vocabulary $V$, running within time
$c\cdot n^\ell$.
For simplicity, we'll use $c\cdot n^\ell$ as common bound on the
iteration of every loop in $P$.
$P^*$ is defined by recurrence on the loop-nesting depth of $P$.
$P^*$ is $P$ if $P$ is loop-free; $(Q;R)^*$ is $Q^*;R^*$;
and $(\bxbf{if}[G]\{Q\}\{R\})^*$ is
$\bxbf{if}[G]\{Q^*\}\{R^*\}$.

If $P$ is $\bxbf{do}[G]\{Q\}$, let $n_0 < \ldots < n_{k-1}$ 
be the ranks in $Q^*$.
Note that we can defined a ``clock" program $C$ that yields for an input 
structure \grs\ 
a listing of size $c \cdot |\grs|^\ell$. Indeed,
by \S \ref{subsec:enumerator} there is an \str-program $E$ that augment any 
$V$-structures
with an enumerator $(\tta_0,\tte_0)$. By \S\ \ref{subsec:arith}
there is a program $M$ that
for a listing $\grn$ as input outputs a listing
of length $c \cdot |\grn|^\ell$.
By \S \ref{subsec:closure} we can compose $E$ and $M$
to obtain our \str-program $C$, generating a listing ($(\tta,\tte)$ of 
length  $c \cdot |\grn|^\ell$.
Choose $C$ with fresh identifiers, and with \tte\
dominating all ranked identifiers in $Q^*$.

Now define $P^*$ to be  \quad
$
C; \; \; \bxbf{do}[G][\tte]\{ \ttb \sdownarrow \tte\tta; 
		\; \myred{\tte\tta \suparrow};
		\; \tta \suparrow;
		\; \tta \sdownarrow \ttb;
		Q^*
$
Since \tte\ dominates all ranked identifiers in $Q^*$, the operation
of $Q^*$ is the same in $P^*$ as in $P$. Also, since the size of \tte\
exceeds the number of iterations of $Q^*$ in $P$, and
the variant \tte\ is contracted in each pass,
in \str\ of the loop above remains the same as in \st.  \qed

\Section{Conclusion and Directions}
The quest for a programming language for PTime has no final destination,
because no language can be both sound and complete for
PTime algorithms. Over the decades a good number of methods were
proposed that were sound and {\em extensionally} complete for PTime,
i.e.\ complete for PTime computabiliTY. But the existence of such
methods is trivial, and the methods proposed so far all miss important
classes of PTime algorithms.  We propose here a novel approach, which
yields a natural programming language for PTime, which is generic
for both inductive data and classes of finite structures, and which
accommodates a substantially broader class of algorithms than 
previous approaches.

We built on \cite{Leivant19,LeivantM19}, where
finite partial-functions form the basic data, and are used as 
loop-variants whose depletion is an abstract form of recurrence.
We consider here a ramification of data that
applied simultaneously to each variant and to its entire rank.
This leads to a programming language \str\ for PTime, which is more 
inclusive than previously proposed works. 

While the purely functional approach of ramified recurrence
does not require a change of semantics of the underlying recurrence
operation, this is no longer the case for our flexible
imperative programming.  The semantics of loops
is modified here to ensure the necessary forms of data depletion, which
in the functional realm are guaranteed by the simplicity of the syntax.
This trade-off is necessary, if we strive for more
algorithmically inclusive programming languages.  The static analysis method,
mentioned in the Introduction, can be called upon to complement the
ICC framework to demonstrate that certain \str\ programs satisfy the
depletion conditions under the standard semantics of looping,
following the line of research of \cite{Kristiansen01,KristiansenN04,
BenamramJK08,Benamram10,BenamramH19} for Meyer-Ritchie's loop programs,
but here with far greater generality.


\newpage

\small

\bibliographystyle{plain}
\bibliography{archiv-feb20}

\normalsize

\newpage

\noindent
{\large\pfont Appendix 1: Proof of Theorem 1.}

\bigskip

The decision problem that asks whether 
a Turing acceptor $M$ fails to accept the empty string is well-known to 
be non-semi-decidable.
We reduce it to $L^p$, thereby showing that $L^p$ is not
semi-decidable either.  Fix a Turing acceptor $F$ running in time $O(n)$,
and a Turing acceptor $N$ running in time $\grW(2^n)$.
Our reduction maps a given Turing machine $M$ to the machine $M'$ that 
on input $x$ simulates the computation of $F$ on input $x$ and, in lockstep,
the computation of $M$ on input \gre.
If the former terminates first,
$M'$ accepts $x$ if $F$ accepts $x$.
If the latter computation terminates first, then
$M'$ switches to simulating $N$ on input $x$.  Thus, if $M$ fails to
accept \gre\ then $M'$ runs in time $O(n)$, and is thus in $L^p$;
whereas if $M$ accepts \gre, say in $k$ steps,
then, since $N$ runs in time $\grW(2^n)$, $M'$ also runs in time
$\grW(2^n)$, with the possible exception of a finite number of inputs
(accepted by $N$ within fewer than $k$ steps). Thus
$M'$ runs in time $\grW(2^n)$, and is thus not in $L^p$, completing
the reduction.
\qed

\newpage

\noindent
{\large\pfont Appendix 2: Examples of fp-structures.}

\medskip

Here are the free structures
for the natural number 3 (i.e.\ the term \mybrown{\bxtt{sssz}}),
the binary string 110 (the term \mybrown{\bxtt{110e}}),
and the binary trees for the terms \mybrown{\bxtt{p(prr)r}}
and  \mybrown{\bxtt{p(prr)(prr)}}.
They use 4,4, 3 and 3 atoms, respectively.
(The vocabulary identifiers are in green, the atoms are indicated by bullets,
and the formal terms they represent are in smaller font.)

\medskip

\xfigtwo{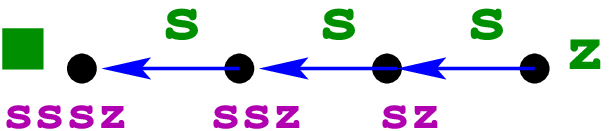}{7mm}{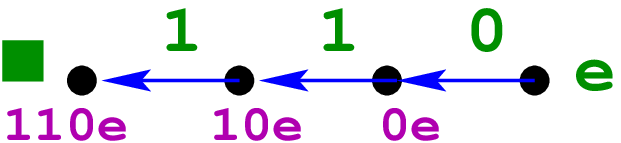}{8mm}

\medskip

\xfig{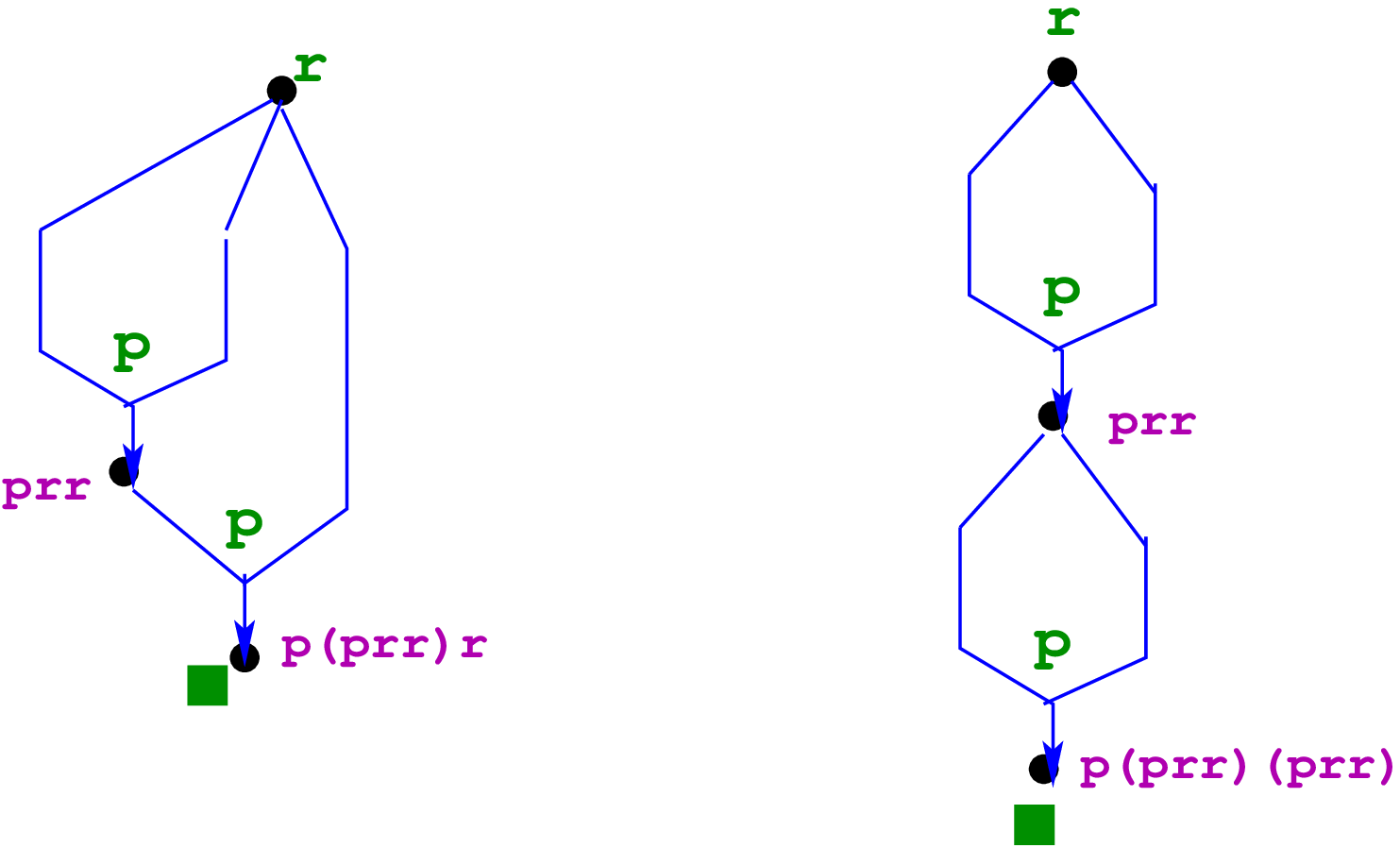}{4.3cm}

\end{document}